\documentclass[superscriptaddress,pre,reprint,showpacs]{revtex4-1}

\usepackage{amsmath,empheq, array, bigints}
\usepackage{bm}
\usepackage{amsfonts}
\usepackage{amsthm}
\usepackage{amssymb}
\usepackage{graphicx}
\usepackage{hyperref}
\usepackage{float}
\usepackage{color}
\usepackage{bm}
\newcommand{\pp}{\bm{p}}
\newcommand{\qq}{\bm{q}}

\definecolor{bluecolor}{rgb}{0,0.,1.}

\definecolor{redcolor}{rgb}{.7,0.,0.}

\usepackage{soul}

\newcommand{\ncd}{\newcommand}
\ncd{\mrm}    {\mathrm}
\ncd{\beq} {\begin{equation}}
\ncd{\eeq} {\end{equation}}

\begin{document}

       \title{Scaling laws and dynamics of hashtags on Twitter}

	\date{\today}

	\author{Hongjia H. Chen}
        \affiliation{School of Mathematics and Statistics, University of Sydney, 2006, NSW, Sydney, Australia}
        \affiliation{Department of Mathematics, University of Auckland, 1010, Auckland, New Zealand}

	\author{Tristram J. Alexander}
        \affiliation{School of Physics, University of Sydney, 2006, NSW, Sydney, Australia}

        \author{Diego F.M. Oliveira}
        \affiliation{U.S Army Research Laboratory, 2800 Powder Mill Rd., Adelphi, MD 20783 USA.}
        \affiliation{Network Science and Technology Center, Rensselaer Polytechnic Institute, 335 Materials Research Center 110 8th St. Troy, NY 12180 USA.}

	\author{Eduardo G. Altmann}
        \email{eduardo.altmann@sydney.edu.au}
        \affiliation{School of Mathematics and Statistics, University of Sydney, 2006, NSW, Sydney, Australia}

	\begin{abstract}
In this paper we quantify the statistical properties and dynamics of the frequency of hashtag use on Twitter. Hashtags are special words used in social media to attract attention and to organize content. Looking at the collection of all hashtags used in a period of time, we identify the scaling laws underpinning the hashtag frequency distribution (Zipf's law), the number of unique hashtags as a function of sample size (Heaps' law), and the fluctuations around expected values (Taylor's law). While these scaling laws appear to be universal, in the sense that similar exponents are observed irrespective of when the sample is gathered, the volume and nature of the hashtags depends strongly on time, with the appearance of bursts at the minute scale, fat-tailed noise, and long-range correlations. We quantify this dynamics by computing the Jensen-Shannon divergence between hashtag distributions obtained $\tau$ times apart and we find that the speed of change decays roughly as $1/\tau$. Our findings are based on the analysis of 3.5 billion hashtags used between 2015 and 2016.
\end{abstract}

\maketitle

\noindent{\bf
  The mathematical study of social systems is only possible because similar processes exist in seemingly different social configurations. Two examples from dynamical systems are rich-get-richer processes -- responsible for the appearance of fat-tailed distributions -- and evolutionary processes -- controlling the dynamics of {\em memes}. Data from the microblogging platform Twitter allow us to study these two generic processes with an unprecedented quantitative accuracy. Here we view hashtags as memes and quantify emerging properties of the collective interaction between these memes, including the appearance of scaling laws and the different time scales involved in their dynamics. 
  }

\section{Introduction}

Hashtags (``\#'') have proven to be one of the most successful innovations in social-media language. They were originally introduced on Twitter to identify topical content in tweets \cite{Hurlock2011}, essentially serving as topic markers to facilitate search and retrieval \cite{ZappavignaSS2015} in the face of an overwhelming amount of information.  For instance, the hashtag ``\#DynamicsOfSocialSystems'' could be used in social-media messages to help users identify comments and papers relevant to this topic. In parallel to this, hashtags also provide a means for users to enhance social ties \cite{ZappavignaSS2015} and conduct a metacommentary distinct from other tweet content \cite{Zappavigna2018}.  Users exposed to a hashtag are invited to use (or modify) the hashtag, starting an imitation~\cite{bagrow2018} and mutation process that leads to a fat-tailed distribution~\cite{mitzenmacher2004,newman2005} of hashtag frequencies \cite{Cunha2011} and that is typical of evolutionary dynamics observed more generally (e.g., in language and in {\em memes}) \cite{NaamanJASIST2011,BeskowIPM2020}.  Hashtags are thus convenient -- can be easily identified and traced -- and generic -- show behaviour seen in various systems (e.g., language, social media, etc.) -- creating thus an ideal scenario for a data-driven study of the dynamics of social systems. 

Previous works examining the dynamical processes underpinning hashtag use have focused on the role of the connections between users on the resulting dynamics \cite{Cunha2011,Romero2011}.  As such these works form part of a more general area of research exploring the nature of user driven dynamics on social media~\cite{gleeson2016,domenico2019}. For instance, models of user behaviour have been able to explain the appearance of a fat-tailed distribution in the distribution of tweets~\cite{refweng,lerman2012social,gleeson2016,notarmuzi2018analytical} and bursty behaviour in the attention of specific topics in Twitter~\cite{domenico2019}.  Other works have focused on specific hashtag dynamics, for instance on the response to an external event \cite{TremayneSMS2014}, for the purposes of ease of analysis while developing data-mining methods \cite{Kapanova2019} or while studying the competition behind diffusion processes \cite{refbingol,ratkiewicz2010characterizing,OliveiraChan}.
Instead, here we are interested not in the dynamics of specific hashtags, but rather in the general statistical behavior of all hashtags used during a particular time window.
By looking at all hashtags simultaneously we account for interactions between different hashtags and we provide an overall statistical characterization of the dynamics of hashtag usage. This is done by repeating classical analyses done in quantitative linguistics for word frequencies~\cite{ferrericancho2001,zanette2005,gerlach2013,fontclos2013,gerlach2014,altmann2016,tanaka-ishii2019}. This approach is justified not only because hashtags can be seen as special types  of words but also because similar dynamical (evolutionary) processes affect the frequency of word usages (albeit at different scales).

The main findings of our manuscript are that hashtags follow statistical laws similar to the linguistic laws observed for words --- such as Zipf's and Heaps' laws -- but that differences appear due to the dynamics of the hashtags. We identify two main aspects of the dynamics of hashtags which differ from natural language: (i) extremely bursty behaviour in the usage of hashtags over time leads to larger than expected fluctuations around the statistical laws, as characterized by an unusual scaling exponent of Taylor's law; and (ii) hashtag usage evolves rapidly with time $\tau$.  We quantify the latter using the (generalized) Jensen-Shannon distance between hashtag observations separated by time $\tau$~\cite{gerlach2016}, and we find a scaling law which characterizes the change in hashtag usage as a function of $\tau$.

This paper is divided as follows. In Sec.~\ref{sec.2} we describe our data and we show relevant time scales of the dynamics. In Secs.~\ref{sec.3} and~\ref{sec.4} we focus on the distribution and scaling behaviour of hashtag frequencies, comparing them to results for word frequencies. In Sec.~\ref{sec.5} we investigate how fast the hashtag distributions change, reporting a new scaling law for the dynamics of hashtags.

\section{Time series of Types and Tokens}\label{sec.2}

\begin{figure}[!bt]
  \centering
  \includegraphics[width=1.1\columnwidth]{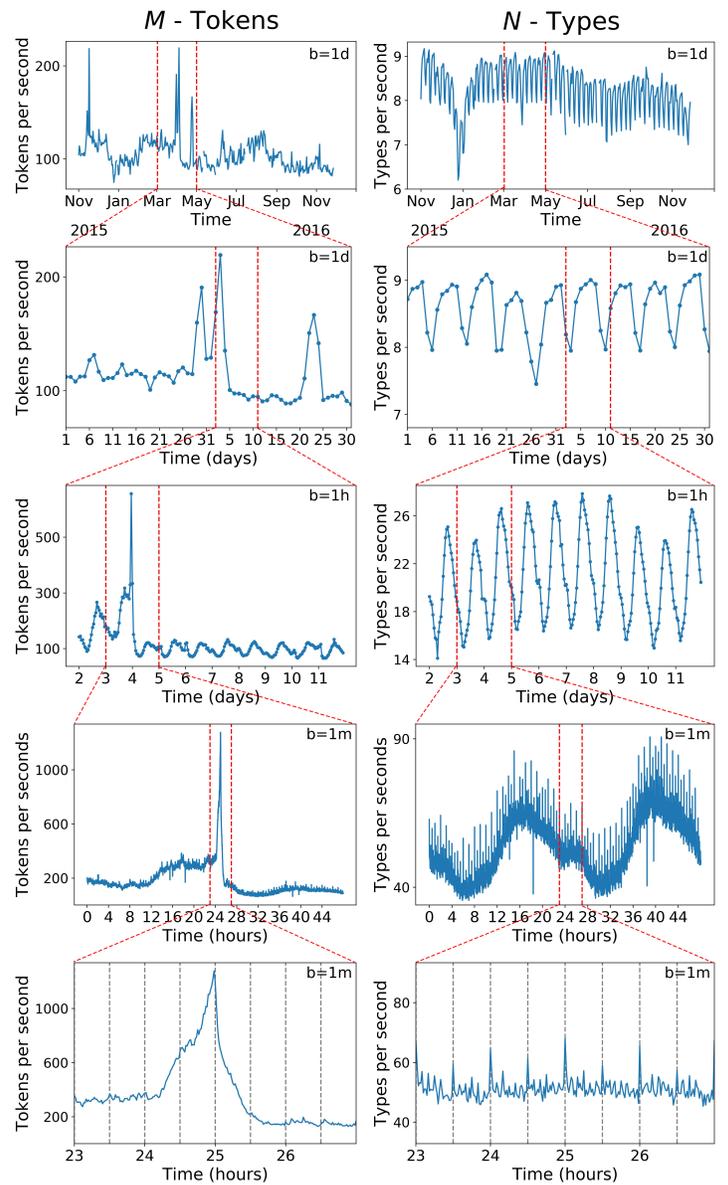}
  \caption{Time series of hashtag tokens $M(t)$  (left column) and types $N(t)$ (right column). Reported (y-axis) is the rate of usage ($N/b$ and $M/b$ with $b$ measured in seconds). Each panel corresponds to a magnification in the time scale (x-axis) of the panel immediately above it in the region indicated by vertical dashed (red) lines.  In the two top panels the data was aggregated at different scales $b$:  $b=1$ day for the two top panels, $b=1$ hour for the middle panels, and $b=1$ minute for the lower two panels~\cite{footnote1}. Time corresponds to GMT.}
\label{fig.1}
\end{figure}

\begin{figure}[!h]
  \centering
  \includegraphics[width=1\columnwidth]{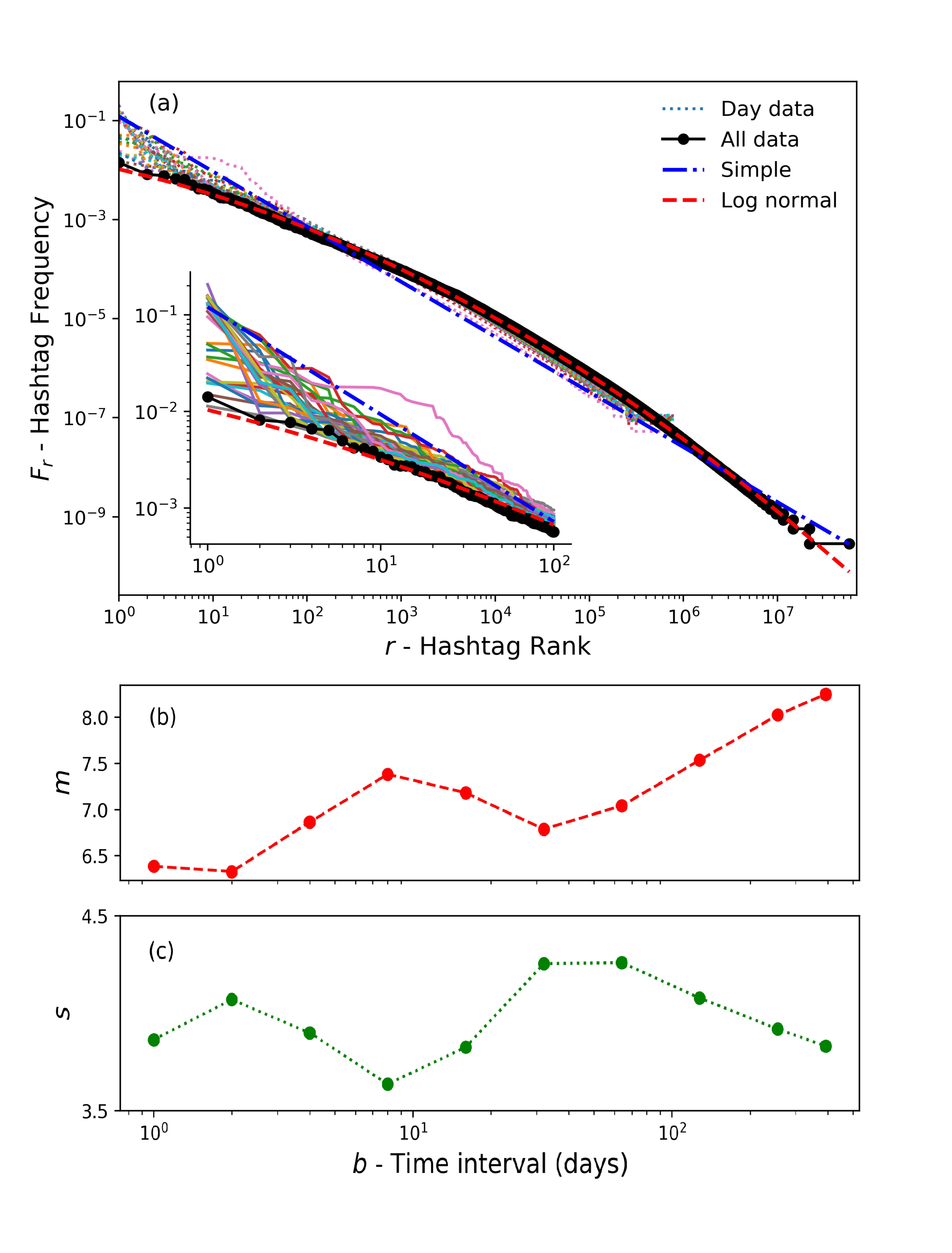}
\caption{Fat-tailed distribution of hashtag frequencies. The solid (black) curve shows the frequency $F_r$ (y-axis) of the $r$-th (x-axis) most frequent hashtag in our complete database $b=392$ days. The thin (colored) lines show the results obtained for $30$ different cases with $b=1$ day.  The inset shows a magnification for small $r$. In the main panel, the dotted line corresponds to Zipf's law~(\ref{eq.zipf}) with the maximum-likelihood parameter $\hat{\gamma}=1.13$ inferred from the data and the dashed (red) line correspond to the best generalized Zipf's law given by Eq.~(\ref{eq.lognormal}) with estimated parameters $\hat{m}=8.25, \hat{s}=3.83$. The lower panels show how the parameters $\hat{m}$ and $\hat{s}$ vary with the size of the database $b$ used in the estimate, i.e., we used all hashtags in the time interval $[t,t+b]$ with $t$ fixed (the minimum) and varying $b$. }
\label{fig.2}
\end{figure}

Our database consists of all hashtags used in a 10\% sample of all tweets published between November 1st 2015 and November 30th 2016. For a given time interval $([t, t + b])$ around time t and of (bin) size $b$, we count how many hashtags were used in our database. Here it is important to distinguish between hashtag types (i.e., unique hashtags) and hashtag tokens (i.e., the repetitive usage of potentially the same hashtags). For instance, in our complete database ($b=392$ days) the hashtag type ``\#mtvstars''  is the most frequently used hashtag (rank $r=1$), responsible for the appearance of $M_{r=1}=49M$ hashtag tokens. Next we have ``\#kca" with $M_{r=2} = 28 M$ and ``\#iheartawards" with $M_{r=3}= 26M$.
Overall, we have $N = 57,876,308$ types and $M = \sum_{r=1}^N M_r = 3,492,300,357$ tokens in our database. We denote $N(t)$ and $M(t)$ as the number of hashtag types and tokens, respectively, in an interval of size $b$ starting at time $t$ \cite{footnote1}.

Figure~\ref{fig.1} shows how the number of hashtag tokens $M$ and types $N$ change in time~$t$ at different time scales. The time series of tokens $M(t)$ shows a more noisy behaviour than the time series $N(t)$ of types: $M(t)$ shows pronounced bursts and spikes while $N(t)$ reflects more clearly the weekly and daily oscillations of Twitter usage. We see a weekly minimum in activity on a Sunday, while the daily maximum occurs around 1600 GMT.  At short time scales, both time series have peaks at the first minute of each hour and each half hour, suggesting that a large number of pre-programmed tweets are being launched at regular patterns. The main peak in $M(t)$ highlighted in this figure is mostly due to the hashtag ``\#iheartawards" which was used during a music awards show that took place in the USA on the 3rd of April 2016 and has rank $r=3$ in our complete database.

\section{Zipf's law}\label{sec.3}

We are interested in the share of total hashtag tokens obtained by the different hashtag types, which can be interpreted as the success rate of individual memes in attracting the attention of users~\cite{Tsur2012}. This possibility of the `rich-getting-richer' element of hashtag use suggests that a fat-tailed distribution should be expected, because of the ubiquity of such a distribution type in data from natural and social systems~\cite{Cunha2011,mitzenmacher2004,newman2005}. Possibly the best known example of such a distribution is Zipf's law, which states that the frequency $F_r = M_r/M$ (i.e., the fraction of all tokens) of the $r$-th most frequent word (type) decays with $r$ as
\begin{equation}\label{eq.zipf}
F_r \sim r^{-\gamma},
\end{equation}
with $\gamma \gtrapprox 1$. 

In Fig.~\ref{fig.2} we show a representation of the hashtag distribution. We observe that a similar distribution is observed for different time intervals, that the distribution spans many orders of magnitude -- in agreement with the fat-tailed character of Eq.~(\ref{eq.zipf}) --, and that the distribution shows a positive concavity (in the double-logarithmic plot) indicating a faster than Zipfian decay. All these observations have been reported for the frequency of words in a recent analysis of Zipf's law in a large data set (Google n-grams)~\cite{gerlach2013} and are consistent with previous analysis of hashtag frequencies~\cite{Cunha2011}.

The observations above motivate us to consider whether generalizations of Zipf's law proposed to describe word frequencies are also describing hashtag frequencies. We considered the distributions and methodology proposed in Ref.~\cite{gerlach2013} to determine which of the eight parameterizations of $F_r$ best describes our hashtag data. Table~\ref{tab.1} lists the different distributions, the best inferred parameters, and a measure of the agreement between data and the (best) distributions. The results show that the best generalized Zipf's law is obtained by a log-normal fit of the rank distribution:
\begin{equation}\label{eq.lognormal}
F_r = C r^{-1} \text{exp}(-\frac{1}{2} (\text{ln}(r) - m)^{2}/s^{2})
\end{equation}
where $C=C(m, s)$ is a normalization constant and $m,s$ are free parameters such that $m<s^2$.  The restriction in the parameter choice is necessary to ensure that Eq.~(\ref{eq.lognormal}) is monotonically decaying in the integers $r$. This is necessary because, by construction, $F_r$ is monotonically decaying (a log-normal distribution in $F_r$ does not imply that the number of hashtag types with a given frequency $M_i/M$ is also log-normal).  A further indication that the distribution~\ref{eq.lognormal} provides a good description of the data for different times $t$ is the fact that the estimated parameters $m$ and $s$ do not strongly depend on the size of the database $b$ (see lower panel of Fig.~\ref{fig.2}). 
This is a different finding from the one reported for natural language, where a double power-law distribution provided a better fit~\cite{ferrericancho2001,gerlach2013}. Differently from the case of language, in the case of hashtags the double gamma distribution (with 3 free parameters) leads to a smaller likelihood $\mathcal{L}$ (or larger $-\log \mathcal{L}$) than the log-normal. Moreover, the parameters of the double gamma in the hashtag distributions differ from the case of language: while for language the first exponent was $\gamma =1$ (as originally proposed by Zipf), in the case of hashtags the first exponent is $\gamma \approx 0.8 < 1$. Altogether, in comparison to word frequencies, hashtags have a slower initial decay of $F_r$ (i.e., the top ranked hashtags have a more similar frequency) and a faster asymptotic decay of $F_r$ (which is faster than a power-law but slower than an exponential).

\begin{table*}[!bt]
\centering
\small
\setlength\tabcolsep{2pt}
\begin{tabular}{ |c|c|c|c| } 
\hline
Model & $F_r \equiv F(r| \; \Omega)$ & Parameter Estimates & $-\log \mathcal{L}/M$ \\ 
\hline
Simple        & $Cr^{-\gamma}$ & $\gamma$ = 1.11 & 11.544\\ 

Shifted Power Law  & $C(r+a)^{-\gamma}$ &$\gamma$ = 1.25, $a$ = 119.8  & 11.205  \\

Exponential cut off  & $C \text{exp}(-ar)r^{-\gamma}$ & $\gamma$ = 0.96, $a$ = 1.11 & 11.195\\ 

Naranan    & $C \text{exp}(-a/r)r^{-\gamma}$  & $\gamma$ = 1.16, $a$ = 5.0931 &  11.347\\ 

Weibull    & $C \text{exp}(-ar^{-\gamma})r^{\gamma - 1}$    & $\gamma$ = -0.24, $a$ = 4.51 &  12.175 \\ 

Log-normal & $C r^{-1} \text{exp}(-\frac{1}{2} (\text{ln}(r) - m)^{2}/s^{2})$   & $m$ = 8.25, $s$ =  3.83 &  {\bf 11.075} \\ 

Double Power Law & 
$
C 
\begin{cases} 
    r^{-1} &  r \leq a\\
    a^{\gamma-1}r^{-\gamma} & r > a 
\end{cases}
$
& $\gamma$ = 1.57, $a$ = 352288.8 & 11.186  \\

Double Gamma & 
$
C 
\begin{cases} 
    r^{-\gamma_1} &  r \leq a\\
    a^{\gamma_2-\gamma_1}r^{-\gamma_2} & r > a
\end{cases}
$
& $\gamma_1$ = 0.8083, $\gamma_2$ = 1.4079, $a$ = 18145.1 &   11.091\\
\hline
\end{tabular}
\caption{Generalized Zipf's law for hashtag frequencies. Different models for the rank-frequency distribution $F_r\equiv F(r|\Omega)$ were fitted to the empirical distribution $F_r$ using the maximum likelihood methods proposed in Ref.~\cite{gerlach2013}. The parameters $\Omega$ that maximize the likelihood $\mathcal{L}$ are reported together with the negative log-likelihood per token $-\log \mathcal{L}/M$ (at the given parameters). The model with maximum likelihood (minimum $-\log \mathcal{L}$) is the log-normal model.}\label{tab.1}
\end{table*}

\section{Heaps' and Taylor's laws}\label{sec.4}

\begin{figure}[!h]
  \centering
    \includegraphics[width=1\columnwidth]{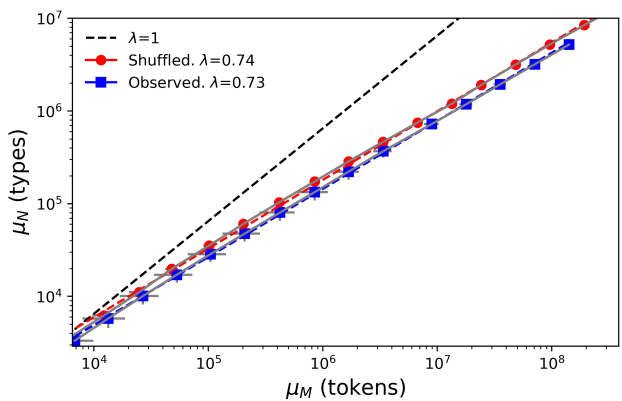}
\caption{The expected number of types~$\mu_N$ and tokens $\mu_M$ scale non-linearly as described by Heaps' law~(\ref{eq.Heaps}). The hashtag data -- $\blacksquare$ with (blue) line -- was obtained using time intervals $b$ ranging from $b=1$ minute to $b=256$ days. The error bars correspond to $\sigma_M$ (x-axis) and $\sigma_N$ (y-axis). The results obtained after shuffling the temporal order of $M(t)$ and $N(t)$ obtained the scale of $b=1$ minute are shown as $\bullet$ with (red) line. The scaling exponents $\lambda$ indicated in the legend  were obtained from a linear regression of the average results (dashed lines).}
\label{fig.3}
\end{figure}

\begin{figure}[!h]
  \centering
    \includegraphics[width=1\columnwidth]{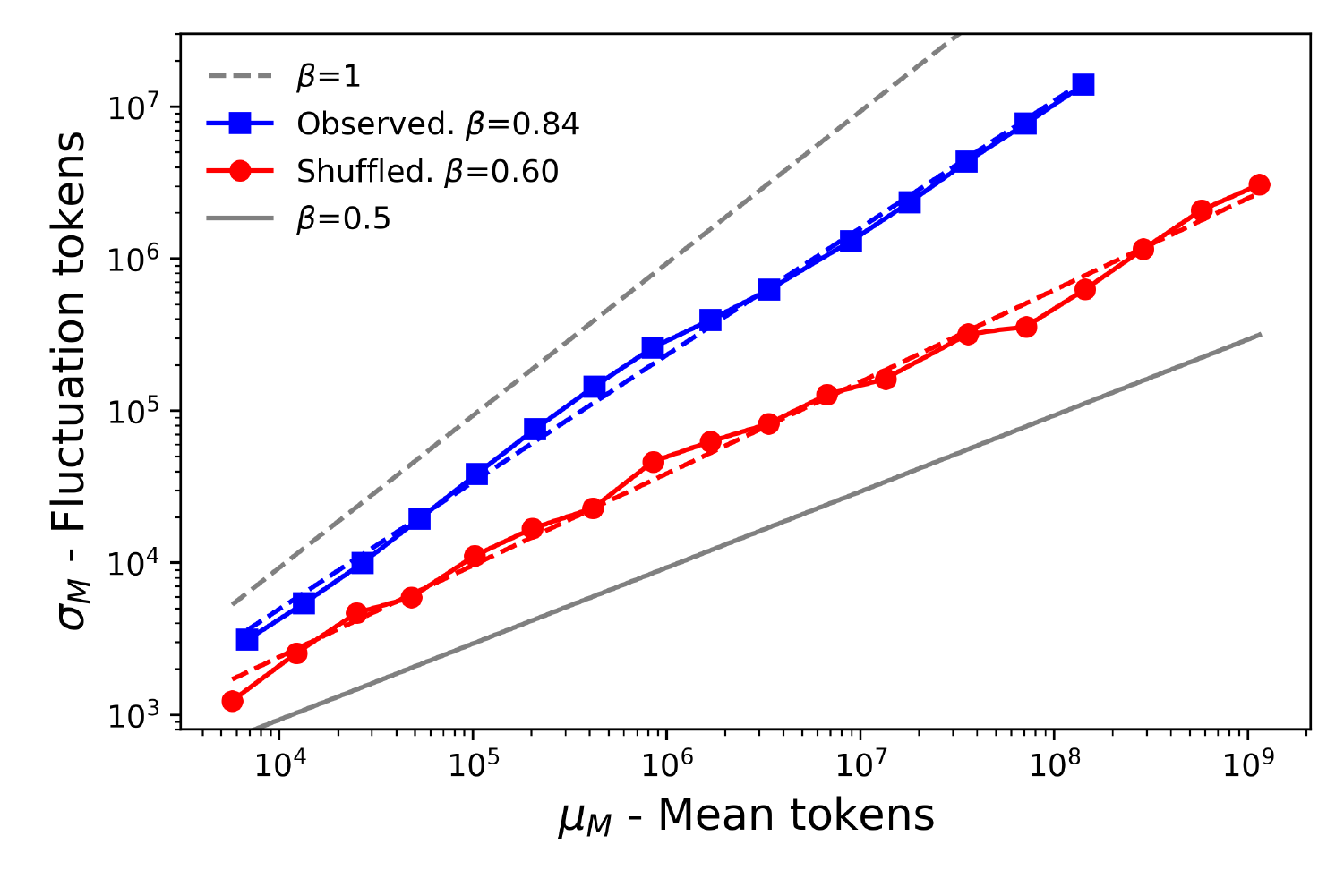}
\caption{Fluctuations and average number of tokens scale non-linearly as described by Taylor's law~(\ref{eq.taylor}). The average $\mu_M$ (x-axis) and standard deviation $\sigma_M$ (y-axis) of the hashtag data -- $\blacksquare$ with (blue) line -- were obtained for intervals ranging from $b=1$ minute to $b=256$ days. The results obtained after shuffling the temporal order of $M(t)$ and $N(t)$ are shown as $\bullet$ with (red) line. The shuffling performed at the scales of $b=1$ minute, $b=1$ hour, and $b=1$ day all showed the same scaling with a different pre-factor. The plot shows a combined curve obtained after re-scaling the curves for $b=1$ hour and $b=1$ day by a constant factor so that they agree with the $b=1$ minute curve. The scaling exponents $\beta$ indicated in the legend were obtained by linear regression (dashed lines).
}
\label{fig.4}
\end{figure}

The Zipfian-type behaviour of hashtag frequencies motivates us to consider also other statistical laws proposed in quantitative linguistics~\cite{ferrericancho2001,zanette2005,fontclos2013,gerlach2013,altmann2016,tanaka-ishii2019}.
We start with Heaps' law, which states that the number of types $N$ and tokens $M$ scale nonlinearly as
\begin{equation}\label{eq.Heaps}
N \sim M^{\lambda},
\end{equation}
where $\lambda <1$ and the symbol $\sim$ indicates that the ratio of the left and right sides tend to a constant for large $M$. To perform this analysis we compute $M(t)$ and $N(t)$ at different time intervals $[t,t+b]$, for different $t$'s and $b$'s as above. We then consider averages $\langle \ldots \rangle$ over all times $t$ for a fixed $b$ and compute the expected value and standard deviation of these quantities as
\begin{eqnarray}
      \mu_M = \langle M(t) \rangle, \sigma_M = \sqrt{\langle M^2(t) \rangle- \langle M(t) \rangle^2}\\
   \mu_N = \langle N(t) \rangle, \sigma_N = \sqrt{\langle N^2(t) \rangle - \langle N(t) \rangle^2}.
\end{eqnarray}
By varying $b$ from minutes to months we effectively vary the size of the database over many orders of magnitude, allowing us to explore the scaling between these quantities.

In our case, Heaps' law~(\ref{eq.Heaps}) is interpreted as the relation between how $\mu_N$ (the expected number of types $N$) scales with $\mu_M$ (the expected number of tokens $M$). The results in Fig.~\ref{fig.3} reveal a striking scaling law over more than four decades, with an estimated exponent $\lambda \approx 0.73$. In this plot we also show the results obtained after shuffling the series at the scale of $b=1$ minute. We observe that $\mu_N$ is increased in the randomized data, reflecting the existing correlation between the hashtags used in neighbouring time intervals~\cite{gerlach2014}. However, the same Heaps law scaling is observed for the shuffled data, in agreement with the previous demonstrations that Heaps' law can be obtained from a random sampling of Zipf's law~\cite{gerlach2014}.

We now investigate how the fluctuations $\sigma$ scale with the mean $\mu$ as
\begin{equation}\label{eq.taylor}
  \sigma \sim \mu^\beta.
  \end{equation}
   Ref.~\cite{reviewtaylor} provides a review of this scaling, known as Taylor's law, showing its appearance and significance in various complex systems. The exponent $\beta=1/2$ is obtained if we consider that the quantity of interest ($M$ in our case) is obtained as the sum of random quantities sampled independently from a distribution with a well-defined second moment. In our case, we can think that the values $M$ in a time interval $[t,t+b]$ is obtained as the sum of the number of hashtags at smaller scales. The case $\beta = 1$ reflects the lack of mixing in the the terms being summed~\cite{reviewtaylor,gerlach2014}. Nontrivial values, $0.5 < \beta < 1$, are obtained in the presence of long-range correlations (in time $t$) or if the underlying distribution from which samples are taken does not have a second moment (large fluctuations of $M$ in small time intervals). In natural language, $\beta=1$ was observed for the case of word types $N$~\cite{gerlach2014} and $0.5 < \beta < 1$ was reported for the fluctuation of individual words~\cite{tanaka-ishii2019}

The results for our hashtag data set are reported in Fig.~\ref{fig.4} and indicate that the exponent $\beta \approx 0.84$ is clearly within the range of non-trivial values (i.e., clearly different from $\beta=0.5$ and $\beta=1.0$). In order to clarify the origin of this non-trivial exponent we repeat the analysis after randomizing the time series $M(t)$. As expected, the exponent after the randomization $\beta_R \approx 0.6$ is smaller than the original exponent. The fact that this exponent is still larger than $1/2$ indicates that the origin of the non-trivial Taylor's law in the hashtag frequencies is due to both long-range correlation in $M(t)$ and sampling from an underlying fat-tailed distribution (with diverging second moment). The latter point is consistent with the bursty behaviour of $N(t)$ reported in Fig.~\ref{fig.1} above, and also with the results of Ref.~\cite{domenico2019}.

\begin{figure}[!ht]
  \centering
    \includegraphics[width=1\columnwidth]{fig5a.pdf}\\\includegraphics[width=0.85\columnwidth]{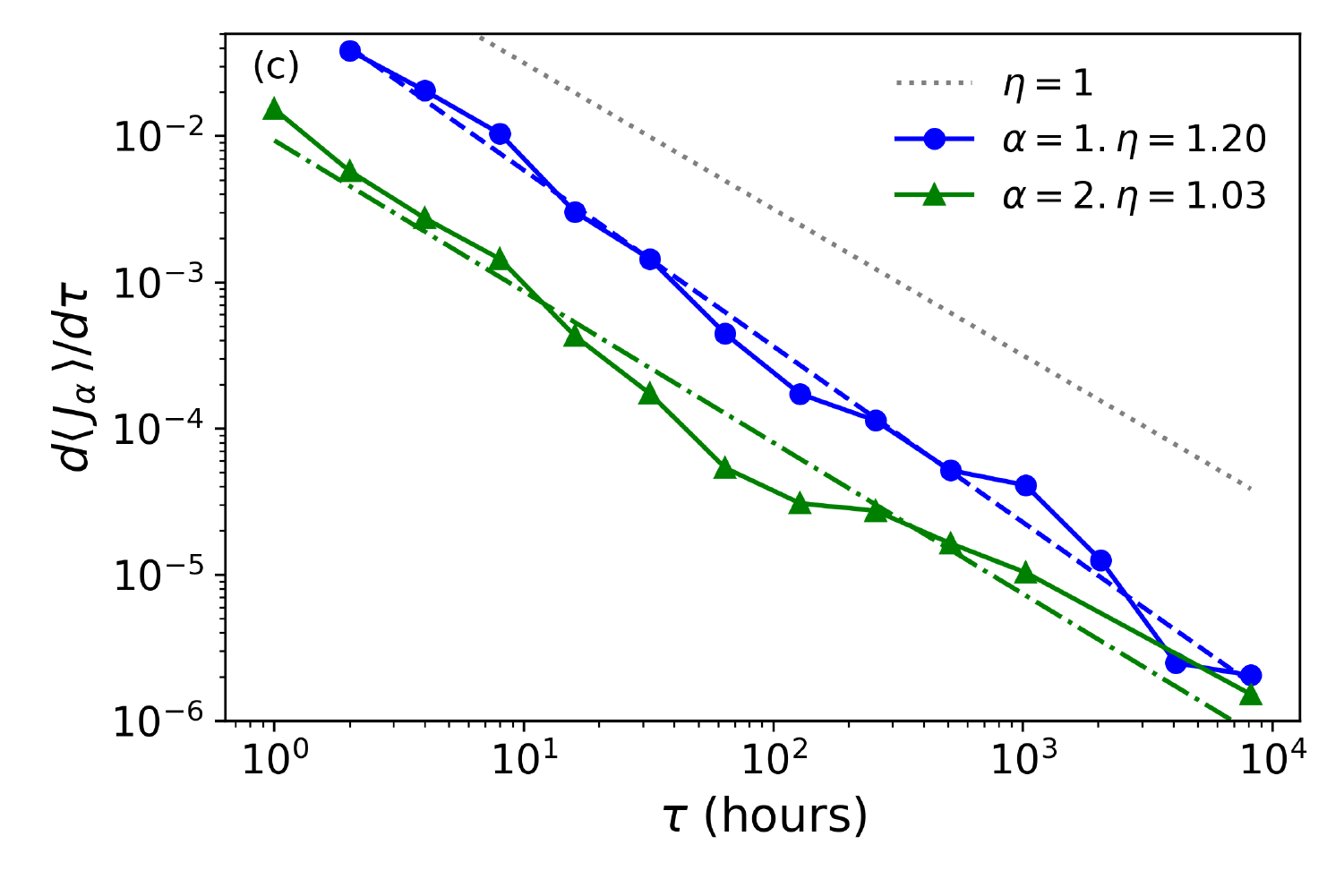}
\caption{Dynamics of hashtags. Panels (a) and (b) show the distance $\langle J_\alpha \rangle$ (y-axis) between the hashtag usage separated by time $\tau$ (x-axis). The symbols (error bars) correspond to the average (standard deviation) $J_\alpha$ computed over all times $t$ for a fixed time separation $\tau$. The $\alpha$ entropy in Eq.~(\ref{eq.alphaH}) was used to compute $J_\alpha$ in Eq.~(\ref{eq.J}) for  $\alpha=1$ (a) and $\alpha=2$ (b). Panel (c) shows the derivative $d\langle J_\alpha \rangle/d\tau$ of the curves in panels (a) and (b), revealing a scaling law between $d\langle J_\alpha \rangle/d\tau$ and $\tau$ The maximum possible value of $J_1$ is $J_1=\sqrt{\ln 2} \approx 0.83$~\cite{gerlach2016}.}
\label{fig.5}
\end{figure}

\section{hashtag Dynamics}\label{sec.5}

\begin{table*}[!ht]
\centering
\small
\setlength\tabcolsep{2pt}
\begin{tabular}{ |c|c|c| } 
\hline
  &  Words in Texts (English) &  Hashtags  in Twitter\\
  \hline
  \hline
  Zipf-like decay of frequency $F_r$       & Yes, faster than $r^{-1}$ & Yes, faster than $r^{-1}$ \\

  Best generalized Zipf's law  &  Double power law~\cite{ferrericancho2001,gerlach2013}
                                 $
                                 \begin{cases} 
    r^{-1} &  r \leq a\\
    a^{\gamma-1}r^{-\gamma} & r > a
\end{cases}
$
                         & Log-normal $C r^{-1} \text{exp}(-\frac{1}{2} (\text{ln}(r) - m)^{2}/s^{2})$ \\
  \hline
  Heaps' law $M \sim N^\beta$  &  $\beta \in [0.52,0.62]$~\cite{gerlach2013} & $\beta=0.73$\\
  \hline
  Taylor's law $\mu \sim \sigma^\lambda$&  $\lambda = 1 $~\cite{gerlach2014} &  $\lambda =0.84$ \\
  \hline
  Dynamics $J_\alpha$ & Linear growth over centuries, constant $dJ/d\tau$ ~\cite{gerlach2016} & Sub-linear growth over months, $dJ/d\tau \sim 1/\tau^\eta$\\
\hline                                                                   
\end{tabular}
\caption{Statistical laws for the frequency of hashtags in Twitter and for the frequency of words in texts.}\label{tab.conclusion}
\end{table*}

So far we have concentrated on general statistical characterizations of hashtag frequencies that remain roughly invariant over time $t$, finding a Zipfian-like  distribution and different scales between the total number of hashtag types $N$ and tokens $M$. Underlying these relationships there is a rich dynamical process of the usage $N_i(t)$ of individual hashtags. Our goal here is to quantify the extent into which, collectively, the frequency of all $M$ hashtags change over time. We use an information theoretic measure to quantify the similarity of two (normalized) frequency distributions, $F_r(t_1)$ and $F_r(t_2)$, following the approach used in Ref.~\cite{gerlach2016} for language.

For each hashtag type $n=1, \ldots, N$ we define the frequency at time $t_s$ as $p_n = M_n(t_s)/M(t_s)$. We consider the frequencies $p_n$ to be an estimate of the probability of using this hashtag and $\pp = (p_1, p_2, \ldots, p_N)$ the probability distribution over all hashtags. The $\alpha$ entropy of $\pp$ is defined as
\begin{equation}
\label{eq.alphaH}
  H_{\alpha}(\pp) = \frac{1}{1-\alpha} \left(\sum_i p_i^{\alpha} - 1 \right)
\end{equation}
and the similarity between two distributions, $\pp$ and $\qq$, can be quantified using the $\alpha$-generalized Jensen-Shannon divergence 
\begin{equation}
\label{eq.jsd}
 D_\alpha(\pp,\qq) = H_\alpha\left(\frac{\pp +\qq}{2}\right) - \frac{1}{2} H_\alpha(\pp) - \frac{1}{2}H_\alpha(\qq).
\end{equation}
For $\alpha=1$ we recover the usual Shannon Entropy $ H(\pp) = -\sum_i p_i \log p_i$ and Jensen-Shannon divergence, which can be viewed as a symmetrized Kullback-Leibler divergence.  Finally, we quantify the similarity between distributions by taking the square root of the divergence
\begin{equation}\label{eq.J}
  J_\alpha \equiv \sqrt{D_\alpha}.
\end{equation}
 As $J_\alpha$ has metric properties for $0\le \alpha \le 2$, it is a natural choice to measure distance. We use  $\alpha=1$ and $\alpha=2$ to obtain different perspectives on the dynamics of hashtags: larger values of $\alpha$ give more weight to high-frequency hashtags~\cite{gerlach2016}. Moreover, the statistical estimators of $J_\alpha$ converge very slowly with sample size $M$ for data with Zipfian frequency distribution~\cite{gerlach2016,koplenig2019} and are better for $\alpha=2$ when compared to the usual $\alpha=1$.

The results obtained for our hashtag data are reported in Fig.~\ref{fig.5} and show rich dynamics. The growth of $\langle J_\alpha \rangle$ with $\tau$ indicates that the measures $J_\alpha$~(\ref{eq.J}) are able to quantify the changes in hashtag frequencies we are interested in. Weekly oscillations are clearly visible in $J_1$ but not in $J_2$, indicating that there are a large number of hashtags that are not among the top ranked ones but are used repeatedly in the same day of the week (e.g., ``\#MondayMotivation"). The overall growth of $J_\alpha$ is slowing down with $\tau$ (i.e., the change in hashtag frequencies is larger for smaller $\tau$'s). Our main empirical finding is that this slow down follows an orderly pattern, described by the scaling law
\begin{equation}
d \langle J \rangle / d\tau \sim 1/\tau^\eta,    
\end{equation}
with $\eta \gtrapprox 1$. This suggests that there is no characteristic time scale for the change of hashtag frequencies in Twitter but that instead it slows down in a self-similar fashion.

\section{Conclusions}\label{sec.conclusion}

In summary, we provided a general statistical characterization of the frequency of hashtags on Twitter. We found that the frequency distribution follows a Zipfian pattern with a faster decay than a simple power-law.  We found that this distribution is well described by the log-normal rank-frequency distribution~(\ref{eq.lognormal}). The type-token relationship shows a scaling law characteristic of Heaps' law, with non-trivial large fluctuations around expected values that follow a fluctuation scaling relationship (Taylor's law). These large fluctuations are due to the very noisy dynamics of hashtag tokens, that shows fat-tailed fluctuations and long-range temporal correlations. We also quantified the collective dynamics due to the change in the frequency $p_i$ of individual hashtags $i$ using a generalized Jensen-Shannon divergence. We found that the distance between hashtag distributions separated by time $\tau$ grows with $\tau$, showing weak oscillations (i.e., distributions at the same day of the week are more similar to each other) and that the velocity of the change decays with $\tau$, following a newly discovered scaling law, $1/\tau^\eta$ with $\eta \gtrapprox 1$. 

A comparison of our findings to previous results for the frequency of words in large collections of texts is given in Tab.~\ref{tab.conclusion}. It reveals striking similarities but also notable differences due to the different dynamics of hashtag and word frequencies. While in texts word tokens of the same word type cluster together, this happens in the middle of many high-frequency function words that permeate the texts with a more regular frequency. In contrast, the appearance of a (new) hashtag can trigger a large response of the usage of the same hashtag, leading to much wider fluctuations and correlations. 
In fact, the top-ranked word in English (``the'') remains the same over centuries, showing a frequency $F_1 \approx 5\%$ that varies only slightly (between $4\%$ and $6\%$) over 200 years (in the Google n-gram database). In contrast, the most frequent hashtag not only varies from day to day but also the frequency of the top ranked hashtag can vary dramatically. For instance, on the first day of our data set (01-11-2015) the top ranked hashtag was ``\#pushawardskathniels'' with a frequency of $4.3\%$, while the hashtag ``\#mtvstars'' was ranked 143rd with a frequency of $0.05\%$. Two weeks later, the hashtags ``\#pushawardskathniels'' and ``\#mtvstars'' were ranked 10th and 1st respectively, with frequencies of $0.7\%$ and $10.9\%$.

A number of our statistical observations are similar to observations reported in isolation in earlier work, such as the burstiness of hashtags and high variability between hashtag volumes~\cite{Tsur2012}, the appearance of fat-tailed distributions in the frequency of hashtags~\cite{Cunha2011}, and the steady evolution of social media language with time~\cite{GrieveELL2017}.  With the combined statistical laws articulated here we hope to provide a framework for generative models to be compared with.  Our findings provide statistical results that (modifications of) existing mechanistic models of social dynamics~\cite{OliveiraChan,bagrow2018}, language~\cite{zanette2005,gerlach2013}, and Twitter~\cite{gleeson2016,domenico2019} should reproduce.  Next steps could be to verify in which extent previous models are able to reproduce our observations and to look in more detail at the nature of the hashtag evolution, e.g., to clarify whether certain types (sub-populations) of hashtags lead to different statistical features or whether the nature of hashtag usage changes more broadly at longer timescales. 

\begin{acknowledgements}
We thank Martin Gerlach for sharing the code that was used in the generalized Zipf's law analysis. 
HHC was funded by a Denison fellowship and EGA and TJA were funded by the CTDS-Incubator Scheme grant number G5121, both from The University of Sydney. 
DFMO was supported by ARL through ARO Grant W911NF-16-1-0524. The views and conclusions contained in this document 
are those of the authors and should not be interpreted as representing 
the official policies, either expressed or implied, of the Army Research 
Laboratory or the U.S. Government. The U.S. Government is authorized to 
reproduce and distribute reprints for Government purposes notwithstanding any copyright notation here on. 
\end{acknowledgements}

\paragraph*{Data Availability:} all data is available at \url{https://doi.org/10.5281/zenodo.3673744}


\begin{thebibliography}{10}
\bibitem{footnote1} An important difference between $N(t)$ and $M(t)$ is that $M(t)$ is simply the sum of $M(t)$ of sub-intervals (of smaller size $b$) while for $N(t)$ this is not the case because we are interested in unique hashtags. This leads to a sub-linear relationship between types and tokens, which is investigated below in the context of Heaps' law.
\end{thebibliography}
\end{document}